# Valley photonic crystal waveguides fabricated with CMOS-compatible process

Takuto Yamaguchi[1]*, Hironobu Yoshimi[1], Miyoshi Seki[2], Minoru Ohtsuka[2], Nobuyuki Yokoyama[2], Yasutomo Ota[3], Makoto Okano[2], and Satoshi Iwamoto[1]*

[1]*Research Center for Advanced Science and Technology (RCAST), The University of Tokyo, 4-6-1 Komaba, Meguro, Tokyo 153-8505, Japan*
[2]*National Institute of Advanced Industrial Science and Technology (AIST), 1-1-1 Umezono, Tsukuba, Ibaraki 305-8568, Japan*
[3]*Department of Applied Physics and Physico-Informatics, Keio University, 3-14-1 Hiyoshi, Kohoku, Yokohama 223-8522, Japan*

E-mail: takutoym@iis.u-tokyo.ac.jp, iwamoto@iis.u-tokyo.ac.jp

Valley photonic crystal (VPhC) waveguides have attracted much attention because of their ability to enable robust light propagation against sharp bends. However, their demonstration using a complementary metal-oxide-semiconductor (CMOS)-compatible process suitable for mass production has not yet been reported at the telecom wavelengths. Here, by tailoring the photomask to suppress the optical proximity effect, VPhC patterns comprising equilateral triangular holes were successfully fabricated using photolithography. We optically characterized the fabricated VPhC devices using microscopic optics with near-infrared imaging. For comparison, we also fabricated and characterized line-defect W1 PhC waveguides, in which the transmission intensities decreased at some regions within the operating bandwidth when sharp turns were introduced into the waveguide. In contrast, the developed VPhC waveguides can robustly propagate light around the C-band telecommunication wavelengths, even in the presence of sharp bends. Our results highlight the potential of VPhC waveguides as an interconnection technology in silicon topological photonic integrated circuits.

# 1. Introduction

The introduction of topology to photonic band theory has shown considerable potential for the manipulation of light propagation. An important property of topological photonic insulators[1-3] is their ability to support edge states, which are impervious to structural imperfections and irregularities. This remarkable hallmark of topologically protected states is beneficial for miniaturization and integration of optical components, allowing to achieve small footprint photonic integrated circuits. Topological photonics have been investigated in various photonic systems, such as arrays of laser-written waveguides[4] and of ring resonators[5], metamaterials[6], and photonic crystals (PhCs)[7-9], realizing different kinds of photonic topological phases. Among these approaches, valley photonic crystals (VPhCs)[10-16] that emulate quantum valley Hall effects are one of the most promising platforms for on-chip photonic integration applications. VPhC structures can be implemented by breaking the spatial inversion symmetry of PhC structures, resulting in high compatibility with conventional semiconductor processes owing to their design simplicity.[10,13,17-21] Topological edge modes referred to as valley kink states, which are supported at the boundary of two VPhC structures with distinct band topologies, exhibit robust light transport around sharp waveguide bends.[22] VPhC waveguides harnessing the valley kink states, facilitate efficient propagation of light with multiple bends in a small area, thus reducing the footprint of photonic circuits and enabling high-density integration. Furthermore, compared to other topological PhC waveguides based on the photonic analog of quantum spin Hall systems,[8-9] VPhC waveguides allow the edge modes to fall below the light line, reducing optical propagation loss. These advantages make VPhC waveguides highly compatible with on-chip integration and attractive for practical applications such as small-footprint optical interconnects.

Multiple VPhC devices have already been reported within the optical regime, such as slow-light devices,[23-25] routing devices,[18-19,26-27] notch filters,[28] and ring lasers.[29-31] However, these results have only been demonstrated using electron beam lithography (EBL). In recent years, silicon photonics has been deployed as a promising technology for photonic integrated circuits because of its compatibility with mature complementary metal-oxide-semiconductor (CMOS) technology that facilitates mass production at low cost. Photolithographically fabricated VPhC waveguides have already been demonstrated in microwave[32] and terahertz bands,[33-35] but their potential in the optical band, except for CMOS-friendly silicon substrate feasibility[18-22], has yet to be investigated. Further, while equilateral triangle-shaped hole-based designs with sharp corners are likely to be used in slab-type VPhC waveguides to obtain a wider operating bandwidth in the optical band,[17-20,22-25,27,31] the sharp corners of the triangular holes are likely to be rounded in photolithography because the resolution of photolithography is lower than that of EBL.

In this study, we devise a photomask pattern for fabricating triangular holes with less-rounded corners and demonstrate its effectiveness in fabricating VPhC patterns on a silicon-on-insulator (SOI) substrate using photolithography. Subsequently, we experimentally demonstrate that VPhC waveguides fabricated by photolithography exhibit robust light transmission through sharp bends around the C-band telecommunication wavelengths. Our findings will help to realize topological photonic integrated circuits using a CMOS-compatible process.

## 2. Valley photonic crystals and waveguides

VPhCs can be realized by breaking the spatial inversion symmetry of systems composed mainly of honeycomb or triangular lattices. Figure 1(a) shows a schematic of a slab-type VPhC waveguide formed by two types of VPhCs comprising equilateral triangular air holes arranged in a honeycomb lattice. The unit cell of the VPhCs is shown in Fig. 1(b), where any two adjacent holes that face away from each other and have different side lengths (denoted as $L_L$ and $L_S$). When $L_L = L_S$, the system has linear dispersion bands constituting Dirac cones centered at points K and K' in the hexagonal Brillouin zone. On the other hand, when $L_L \neq L_S$, the band degeneracy is lifted owing to the lack of spatial inversion symmetry. The resultant band gaps are formed between the frequency extremes of the bands at the K and K' points (the so-called valleys). The topological invariant of the gapped bands is defined by the valley Chern number,[10] which flips its sign at a valley in two VPhC structures in which the hole sizes $L_L$ and $L_S$ are swapped. Along the interface between structures with distinct valley Chern numbers, such as the zigzag interface in Fig. 1(a), a topologically protected state exists that is backscattering-immune against sharp bends. This valley kink state has potential applications in ultra-compact topological photonic devices. Figure 1(c) shows the calculated TE-like mode dispersion for the zigzag interface of VPhCs on an air-suspended silicon slab with a refractive index of 3.48, a lattice constant $a = 500$ nm, a slab thickness $d = 220$ nm, and equilateral triangular holes of side lengths $L_L = 1.25a/\sqrt{3}$ and $L_S = 0.75a/\sqrt{3}$. According to calculations using the three-dimensional plane wave expansion (3D PWE) method, a valley kink state exists in the bulk band gap below the light line. Although slab-type VPhC waveguides can be designed similarly using circular holes, a design with triangular holes is preferred to obtain a larger bulk bandgap,[8] resulting in a wider operating bandwidth.

## 3. Fabrication

We proposed a special photomask pattern to reproduce the acute angles of triangles as much as possible using optical proximity effect correction (OPC). As shown in the left side of Fig. 2(a), we designed a correction pattern with small kite-shaped polygons whose side lengths are 64 and 110 nm, partially overlapping the corners of the triangular pattern. Device fabrication was conducted at the AIST's 300-mm CMOS R&D foundry facilities, where 193-nm ArF immersion photolithography and dry etching techniques were used to define the device patterns on an SOI substrate. Figure 2(b) shows two scanning electron microscopy (SEM) images of isolated triangular holes fabricated by an exposure dose of 20.5 mJ/cm$^2$ using each of the photomask patterns in Fig. 2(a). Although these two fabricated triangular holes were approximately the same size (the sides of the circumscribed triangles were approximately 400 nm), the tendency for the triangles to be fabricated with rounded corners was suppressed using our correction pattern. If we define the filling ratio as a fraction with the area of the circumscribed triangle as the denominator and the area of the hole as the numerator, the filling ratio was improved from 79.2 to 86.9% before and after considering the OPC.

When fabricating the VPhC structures, we prepared a photomask by applying our correction pattern only to the small triangular holes owing to the limitations of the foundry's design rules and slightly lowered the exposure dose to 20 mJ/cm$^2$. Figure 3(a) shows an enlarged SEM image of the zigzag interface of the fabricated VPhC waveguide, where the unit cells comprising the upper and lower halves of the VPhCs are inverted in the vertical direction. This SEM image indicates that our OPC pattern was effective even for fabricating small triangular holes in the VPhC structure. Using this technique, we fabricated straight and Z-shaped VPhC waveguides formed by a zigzag interface, as shown on the left and at the center panels of Fig. 3(b). The lattice constant *a* of the VPhCs was 460 nm and the waveguide length was 140*a* (64.4 μm). For optical characterization, we fabricated devices in which the VPhC waveguide was butt-coupled to wire waveguides, as shown in Fig. 3(c). The wire waveguides were terminated using a broadband grating coupler designed for optical input and output. For comparison, we also fabricated a device in which a straight VPhC waveguide was replaced with a bulk VPhC structure with no interface (right panel of Fig. 3(b)). The total length of the wire waveguides in a single device was fixed at the same value for all devices. For the optical measurement chips used in the next section, a device layer with a slab thickness of 220 nm was fully embedded using CVD-grown silica cladding. The SEM images shown in Fig. 3 were taken before depositing the silica over-cladding layer.

For comparison with the VPhC devices, we fabricated 140-period-long W1 PhC waveguides with straight and Z-shaped line defects with a lattice constant of 400 nm. The W1 PhC waveguides were formed by removing a single row of circular holes arranged in a triangular lattice. A device with no line defect was also fabricated. The silicon slab thickness for these PhC devices was the same of that of the VPhC devices.

## 4. Measurement methods

Figure 4 shows a schematic diagram of the optical characterization setup. We used a wavelength-tunable laser diode (Santec TSL-510) to obtain laser light in the wavelength range of 1480–1640 nm. After being reflected by two mirrors and passing through a linear polarizer and beam splitter, the laser light was focused on the grating coupler of the fabricated device through an objective lens (OLYMPUS LMPlan IR 100x, NA = 0.80). The sample was placed on a 2D motorized stage to allow precise positioning of the device with respect to the incident light. A linear polarizer was placed such that only the TE-polarized light was incident on the grating coupler at the input. The light emitted from the sample was collected using an objective lens and reflected by beam splitters to an InGaAs near-infrared (NIR) imaging camera (Artray ARTCAM-008TNIR). The transmitted intensity was measured by integrating the signals measured on the pixels within the area of interest (a fixed-size rectangular region, including the output grating coupler) in the captured NIR image. To determine the location of the device during characterization, the irradiated area was simultaneously observed with a charge-coupled device (CCD) camera using a Köhler illuminator with a white light-emitting diode.

## 5. Optical characterization results

### 5.1 Transmission characteristics of W1 PhC devices

Figure 5(a) shows the SEM images of a portion of the fabricated W1 PhC devices before the top silica cladding was deposited, where the diameter of the circular holes was approximately 187 nm. Figure 5(b) shows the optical intensity curves of the W1 PhC devices as functions of the incident light wavelength.

To investigate the experimental results in detail, we acquired the contour data of a circular hole of the W1 PhCs from the SEM observations of the devices before the deposition of the upper $SiO_2$ cladding and calculated the band diagram of our fabricated W1 PhC waveguides using 3D-PWE, as shown in Fig. 5(c). The refractive index of the silica cladding was set to 1.45. According to the band calculation results, the W1 PhC waveguides supported laterally even or odd waveguide modes in the wavelength range of 1498–1622 nm. The operating bandwidth can be divided into three regions, and the transmission spectra exhibit different behaviors in each region. In the wavelength region between 1584 and 1622 nm (Region I in Fig. 5(c)), only the even mode existed below the cladding light line. In this region, the transmission spectrum of the Z-shaped waveguide exhibited sharp dips at certain wavelengths (approximately 1590 and 1606 nm) owing to backscattering at sharp turns. Furthermore, in the slow-light region on the longer-wavelength side than 1612 nm, the signal detected from the Z-shaped waveguide was significantly weaker than that from the straight waveguide. In region II (1558–1584 nm), the even mode was forced above the light line of the silica, making light propagation through sharp turns even more difficult. Finally, even and odd modes coexisted in region III at wavelengths shorter than 1558 nm, resulting in mode conversion between the even and odd modes at sharp bends. Furthermore, the odd mode of the W1 PhC waveguides were hardly coupled with a wire waveguide owing to symmetry mismatch. Because of these influences, in addition to the aforementioned light-line problem, it was estimated that few transmission signals can be detected from a Z-shaped waveguide in this wavelength range. As can be seen from these results, W1 PhC waveguides are unable to guarantee robust light propagation against sharp turns. These trends in the transmission curves of the W1 PhC waveguides are consistent with experimental results for a similar W1 PhC waveguide structure with multiple 60-degree bends.[36]

5.2 Transmission characteristics of VPhC devices

We characterized three types of VPhC devices. Figure 6(a) shows the SEM images of a portion of the fabricated VPhC devices before the top silica cladding was deposited. During fabrication, we designed a photomask without OPC patterns to allow the small triangular holes of the VPhCs located near the edges of the silicon slabs to match the foundry's design rules, resulting in the formation of tiny round holes, as shown on the left and right panels of Fig. 6(a). Compared with the fabrication results of VPhCs in the region away from the slab edges, it can be seen that our correction pattern is effective particularly in fabricating small triangular-shaped holes (roughly 220 nm on a side). Figure 6(b) shows the transmission spectra of the fabricated VPhC devices, where the signal intensity transmitted through the straight VPhC waveguide was much larger than that transmitted through the VPhC bulk at wavelengths longer than approximately 1510 nm. In addition, by comparing the transmission curves of the straight and Z-shaped waveguides, the detected signal intensities were found to be nearly equivalent in the wavelength range of 1510–1575 nm. This result indicates that the VPhC waveguides fabricated by photolithography allowed light to propagate robustly through sharp turns around the telecom C band. Figure 6(c) shows the photonic band diagram of the fabricated VPhC waveguides, calculated in the same manner as shown in Fig. 5(c). The single-mode bandwidth of our VPhC waveguide was calculated in the range of 1505–1572 nm (yellow regions in Figs. 6(a) and 6(b)), which corresponds well to the bandwidth over which light can be transmitted through the Z-shaped VPhC waveguide in our experiment. This indicates that the VPhC devices fabricated by photolithography can realize robust light guiding even in the presence of sharp waveguide turns as VPhC waveguides fabricated with EBL. Interestingly, in this single-mode bandwidth, the guiding mode preserves robustness against sharp turns, even though the mode lies above the light line of the cladding material, which is particularly different from the characterization results of the W1 PhC waveguides (corresponding to region II in Figs. 5(b) and 5(c)). Although the physical mechanisms underlying this phenomenon remain unclear and require further investigation, this unique feature could be another advantage of VPhC waveguides over W1 PhC waveguides.

## 6. Conclusion

In conclusion, we fabricated VPhC devices on a silica-clad SOI wafer by photolithography using a photomask with an OPC pattern devised to fabricate triangular holes with side lengths of several hundred nanometers. The optical characterization of the fabricated devices was performed using microscope optics with an NIR imaging camera. For comparison, we fabricated similar devices with a W1 PhC structure and experimentally confirmed that such a conventional PhC waveguide using line-defect modes is unable to show a continuous transmission spectrum in the presence of sharp waveguide bends. By comparing the spectra of the transmission signals emitted from straight and Z-shaped VPhC waveguides of the same length, it was found that light can propagate smoothly through the sharp bends in our VPhC waveguides in the bandwidth over the C band. In addition, this transmission bandwidth was in close agreement with the operating bandwidth of the VPhC waveguide obtained from photonic-band calculations. These results demonstrate the topologically protected transport properties expected in VPhC waveguides, even for VPhC devices fabricated via photolithography. Our demonstration strengthens the possibility of employing VPhC waveguides as optical interconnects and will be a cornerstone for topological photonic integrated circuits in silicon photonic platforms.

## Acknowledgments

This study was supported by JSPS KAKENHI (Grant No. JP15H05700, JP15H05868, JP17H06138, JP20J22862, and JP22H00298), JST CREST (JPMJCR19T1), and the New Energy and Industrial Technology Development Organization (NEDO).

# Figure Captions

**Fig. 1.** (a) Schematic of a slab-type VPhC waveguide composed of equilateral triangular holes. (b) Rhombic unit cell of the VPhCs, comprised of large and small triangular holes with side lengths of $L_L$ and $L_S$, respectively. (c) Calculated dispersions of a valley kink state in an air-suspended zigzag-type VPhC waveguide on a silicon slab.

**Fig. 2.** (a) Photomask patterns used for fabricating an isolated triangular hole using photolithography. OPC pattern (left) designed with a kite-shaped polygon partially superimposed over the corners of the initial triangle pattern (right). (b) SEM images of isolated triangular holes fabricated based on photomask patterns shown in (a). Green dotted lines represent the contour of the side wall of the fabricated holes. Red lines describe the circumscribed triangle of the hole contours. Lines are drawn by image processing of SEM images.

**Fig. 3.** False-colored top-view SEM images of fabricated VPhC devices before depositing upper cladding onto the silicon device layer. (a) Magnified view of a fabricated VPhC waveguide. (b) Three types of VPhC devices under investigation. (c) Overall view of device for optical characterization. Square dummy patterns added in sparse pattern areas to equalize the pattern density and homogenize process conditions across the wafer.

**Fig. 4.** Schematic of optical setup for near-infrared imaging (TLD: tunable laser diode, M: mirror, LP: linear polarizer, BS: beam splitter, OL: objective lens, MS: motorized stage, NIR: near-infrared camera, CCD: charge-coupled device camera, KI: Köhler illuminator).

**Fig. 5.** (a) Top-view SEM images of a portion of the W1 PhC devices before the CVD deposition of $SiO_2$. (b) Optical intensity spectra obtained from three types of W1 PhC devices. (c) Calculated dispersion relations for guided modes in our fabricated W1 PhC waveguides. Solid red and dotted green lines indicate laterally even and odd TE-like guided modes for W1 PhC waveguide, respectively. Numerical simulation performed by 3D PWE method using contour data of a circular hole obtained from SEM image processing. Black dashed and dotted lines correspond to light lines of air and $SiO_2$, respectively. Green shaded

regimes represent bulk mode regions. Frequency region where the even mode lies within the bulk bandgap in the calculated band diagram is divided into three colored areas (Region I/II/III) in (b) and (c) according to its dispersion characteristics.

**Fig. 6.** (a) Top-view SEM images of a portion of the VPhC devices before the CVD deposition of $SiO_2$. (b) Optical intensity spectra obtained from three types of VPhC devices. (c) Dispersion relations for TE-like guided mode in our fabricated VPhC waveguide. Yellow areas in (b) and (c) highlight of the calculated operating bandwidth where the guided mode lies below the light line of air and above the silica light line.

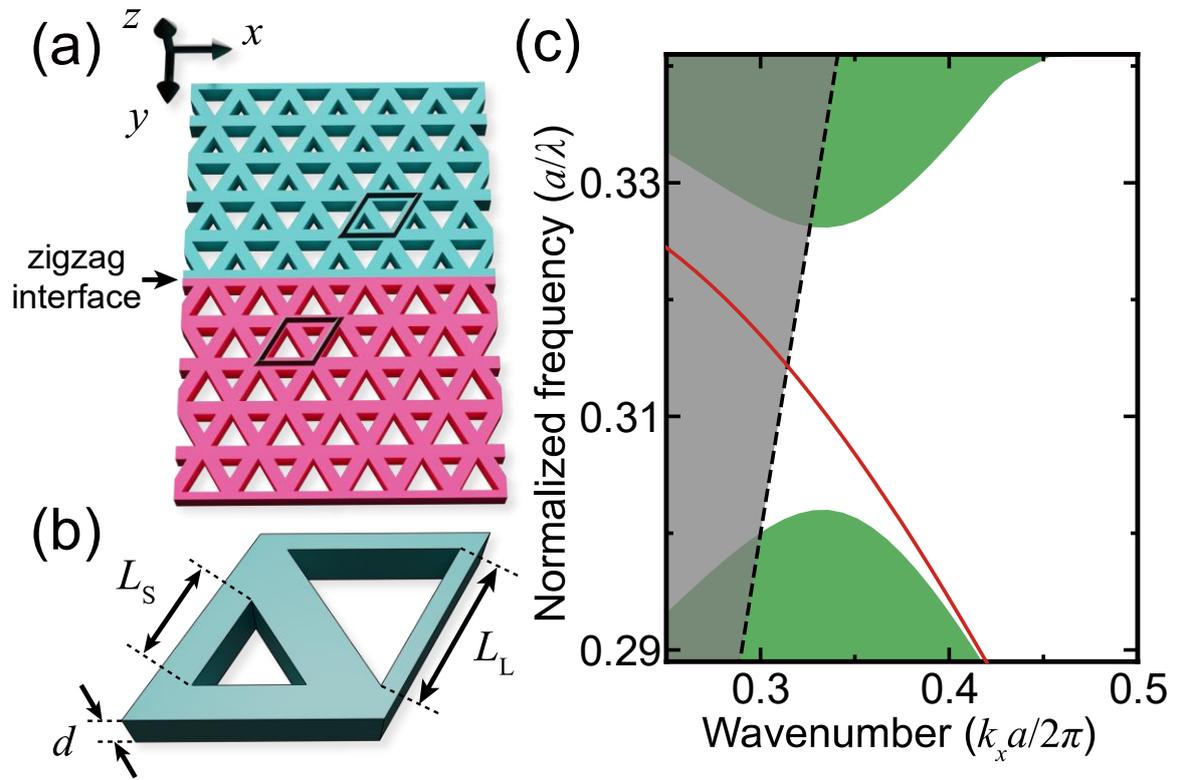

Fig. 1.

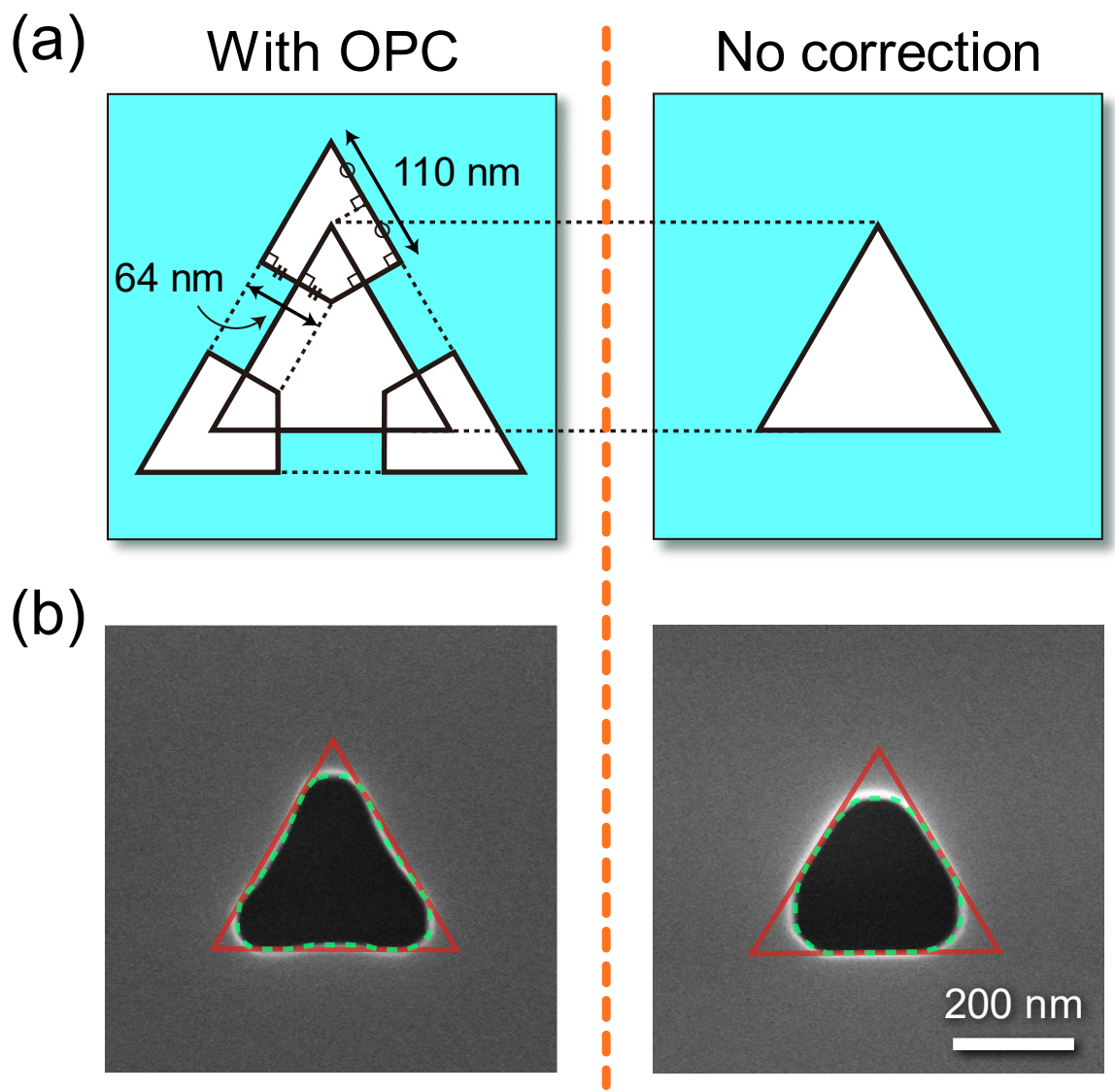

Fig. 2.

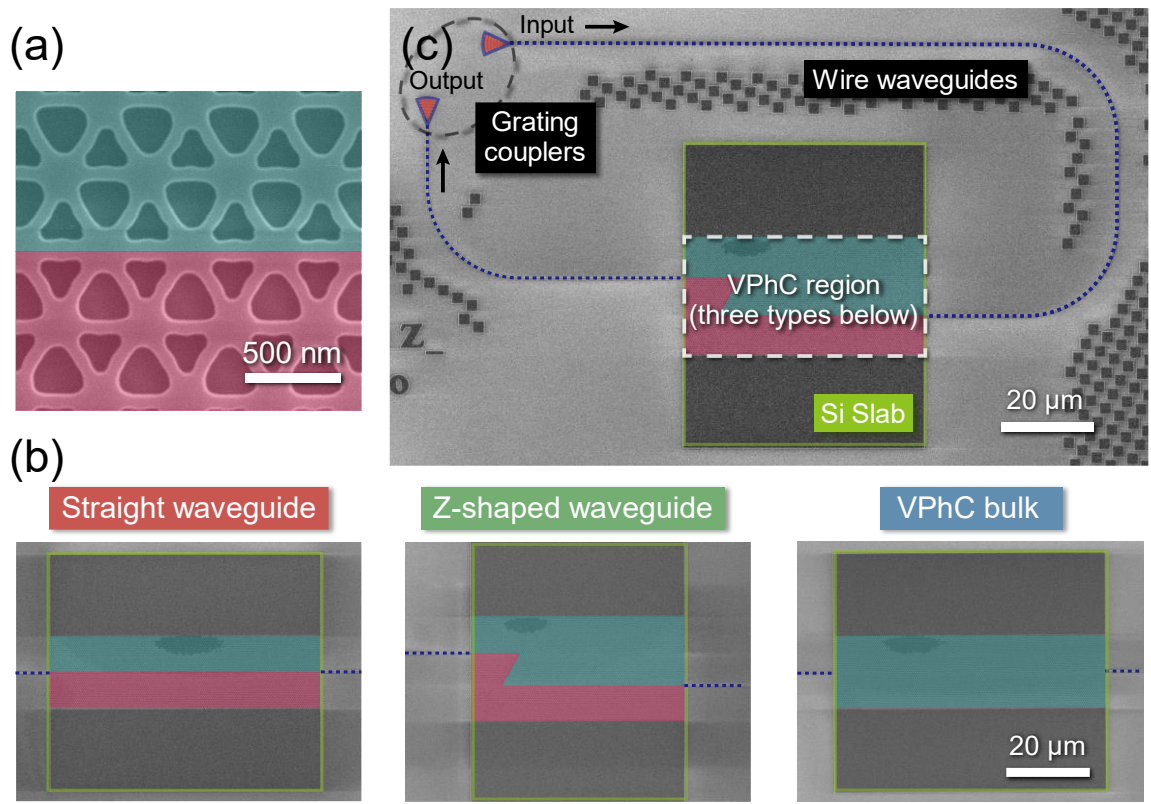

Fig. 3.

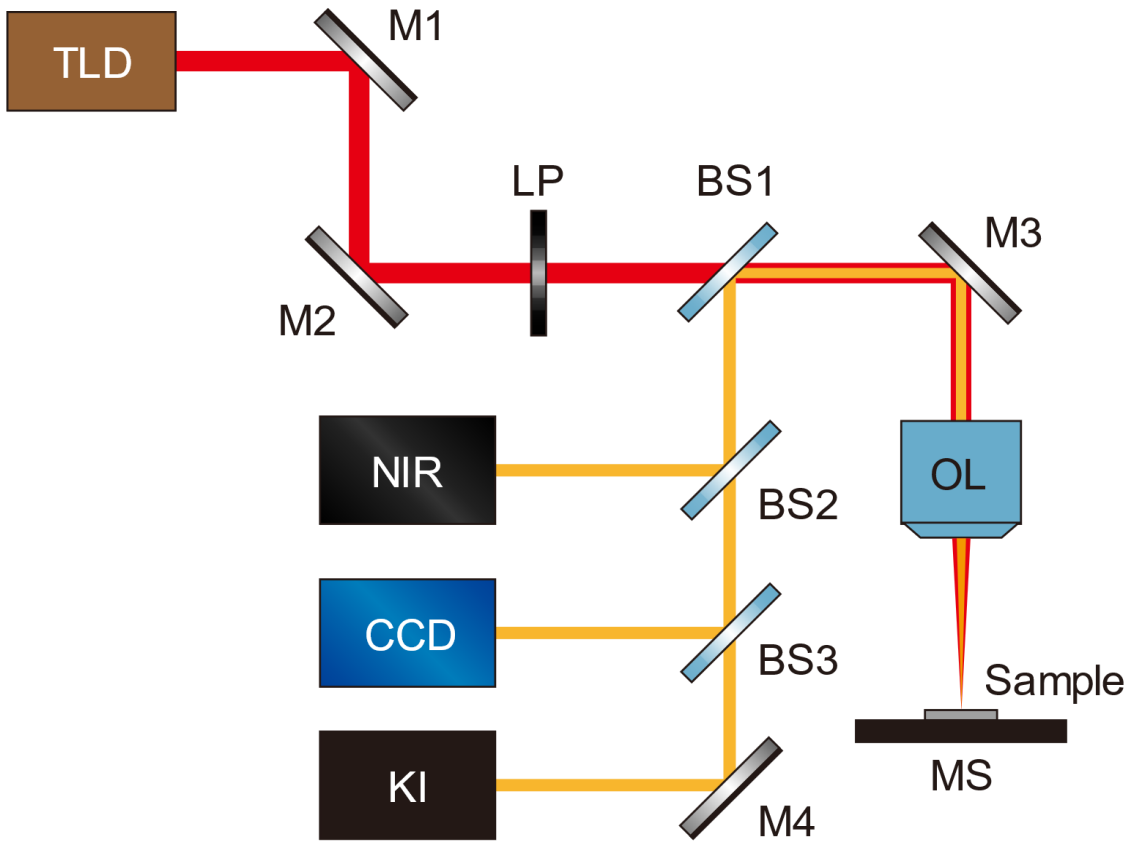

Fig. 4.

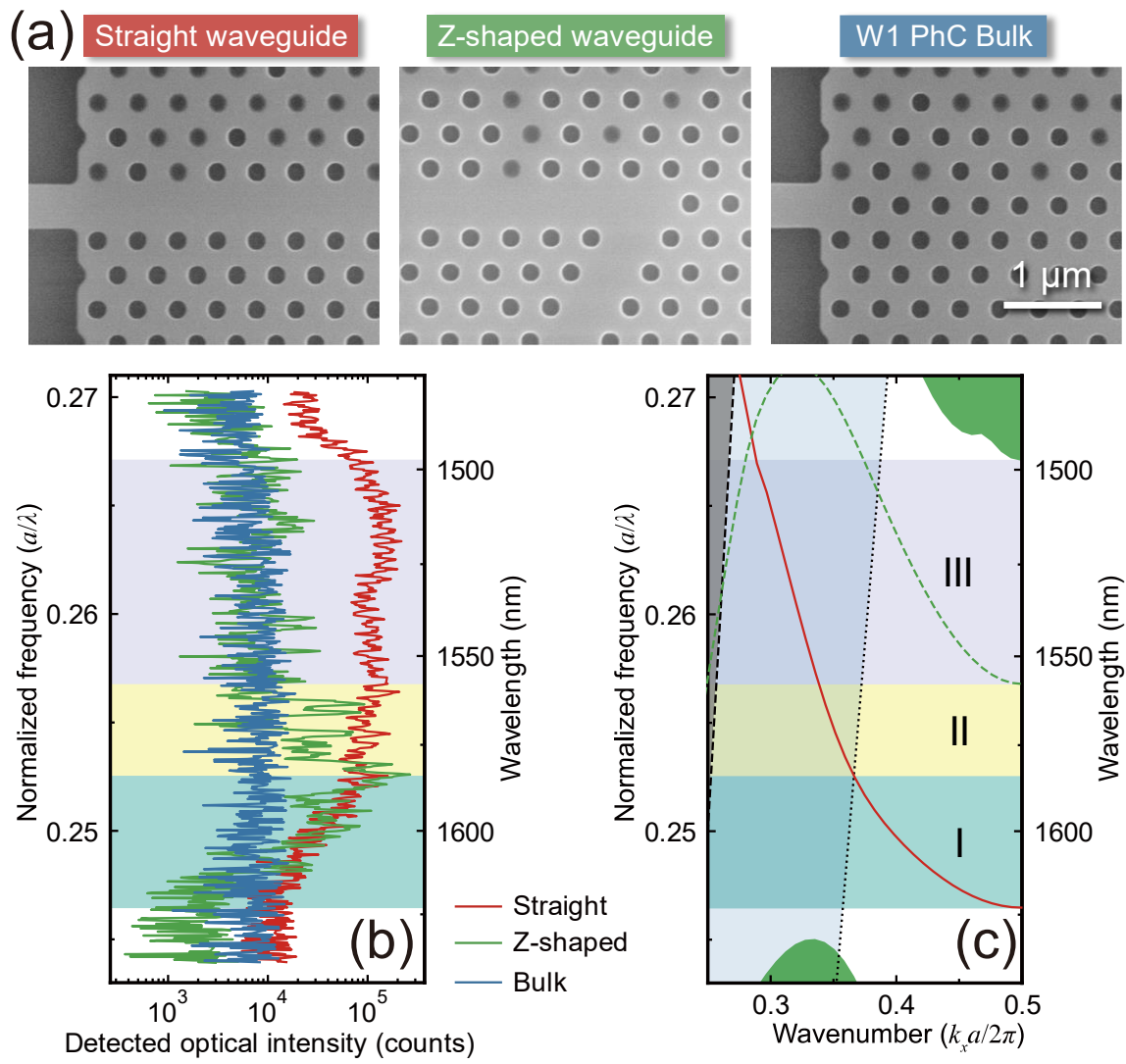

Fig. 5.

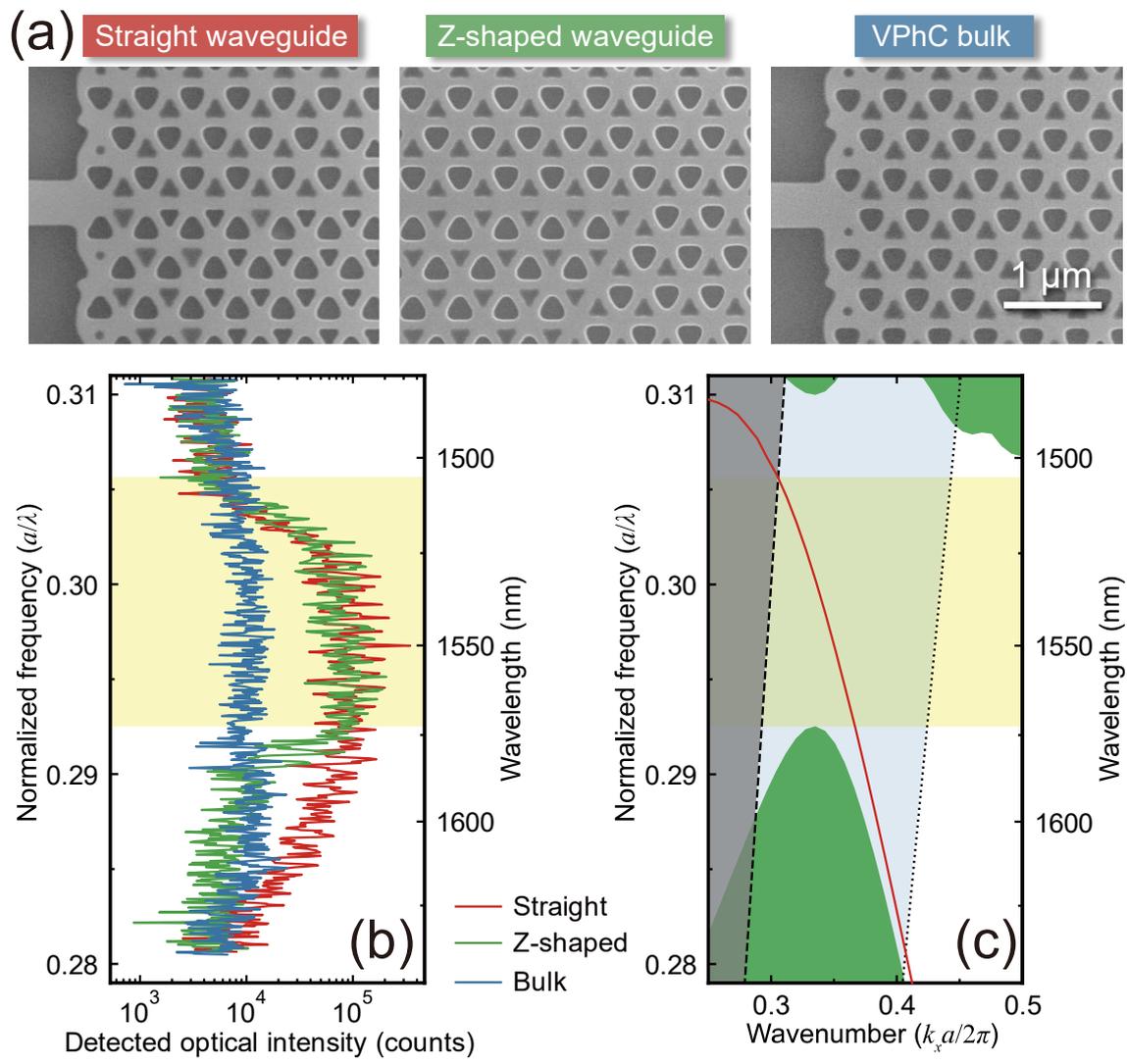

Fig. 6.